\documentclass[
 twocolumn,
 aps,
 prl,
 reprint,
 showpacs
]{revtex4-1}

\usepackage{amsmath}
\usepackage{graphicx}

\usepackage[
 colorlinks=true
 ]{hyperref}

\begin{document}

\title{Overcoming losses with gain in a negative refractive index metamaterial}

\author{Sebastian Wuestner}
\author{Andreas Pusch}
\author{Kosmas L. Tsakmakidis}
\author{Joachim M. Hamm}
\author{Ortwin Hess}
\email[Corresponding author. ]{o.hess@surrey.ac.uk}

\affiliation{Advanced Technology Institute and Department of Physics, University
of Surrey, Guildford, GU2 7XH, Surrey, United Kingdom}

\date{\today}


\begin{abstract}
On the basis of a full-vectorial three-dimensional Maxwell-Bloch approach
we investigate the possibility of using gain to overcome losses in
a negative refractive index fishnet metamaterial. We show that appropriate
placing of optically pumped laser dyes (gain) into the metamaterial
structure results in a frequency band where the nonbianisotropic
metamaterial becomes amplifying. In that region both the real and
the imaginary part of the effective refractive index become simultaneously
negative and the figure of merit diverges at two distinct frequency
points. 
\end{abstract}

\maketitle

Negative refractive index metamaterials offer the possibility of revolutionary
applications, such as subwavelength focusing \cite{Pendry_2000},
invisibility cloaking \cite{Pendry_2009}, and ``trapped rainbow''
stopping of light \cite{Tsakmakidis_2007}. The realization of these
materials has recently advanced from the microwave to the optical
regime \cite{Shalaev_2006,Valentine_2008}. However, at optical wavelengths
metamaterials suffer from high dissipative losses due to the metallic
nature of their constituent meta-molecules. It is therefore not surprising
that overcoming loss restrictions is currently one of the most important
topics in metamaterials research \cite{Zheludev_2010}.

It has been suggested that, owing to causality, simultaneous loss-compensation
and negative refractive index might only be attainable in a very narrow
bandwidth with high losses nearby \cite{Stockman_2007}. In an ongoing
discussion several authors have reasoned that causality-based criteria
have to be applied carefully and do not in general lead to such a strict result
\cite{Skaar_2006,Kinsler_2008}. This said, the theoretical possibility
to compensate losses in optical metamaterials does not necessarily
imply that the gain available from optically active media suffices
to achieve this goal. Indeed, bulk gain coefficients are usually an
order of magnitude smaller than the absorption coefficients of metals
at optical frequencies.

A vital clue as to how the aforementioned limitation could be overcome
came from \cite{Zheludev_2008}. There it was shown that the
incorporation of gain in regions of high field intensity gives rise
to an effective gain coefficient that can exceed its bulk counterpart by potentially orders 
of magnitude. Exploiting this gain-enhancement effect loss reduction and giant field enhancement have been
reported for a double-fishnet metamaterial using frequency-domain models \cite{Sivan_2009,Dong_2009,*Dong_2010}.
The particular model used in \cite{Sivan_2009} accounts for spatial nonuniformity and saturation of the gain.
However, the dynamic evolution of the gain system, nonlinearly pumped
by a short intense pulse, cannot be described self-consistently
with such a frequency-domain approach and relies on assumptions of
cw-excitation. A time-domain calculation of gain in two-dimensional
electric or magnetic metamaterials has recently been reported in \cite{Fang_2009,*Fang_2010},
but therein the effect of the pump field has again not been considered
self-consistently.

In this Letter, we study the optical response of a negative refractive
index (NRI) metamaterial with a gain medium embedded in the structure
and we find that complete loss-compensation and even amplification
is possible using realistic gain parameters. 
To this end we use a full-vectorial time-domain approach that manages 
to self-consistently couple the evolution of the occupation densities in 
the gain medium directly to Maxwell's equations in three dimensions 
\cite{Bohringer_2008,*Bohringer_2008_2,Klaedtke_2004,Klaedtke_2006}.
Nonlinearity, saturation of the gain medium, and spatio-temporal variations
of both absorption and emission are inherent to our model, avoiding
the need for external, precalculated inputs.

\begin{figure}[b]
 \begin{centering}
  \includegraphics{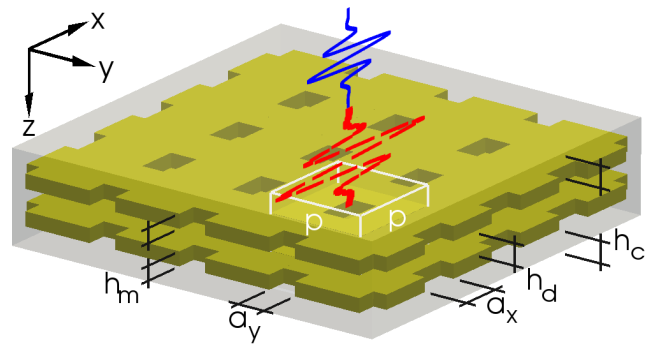} 
 \end{centering}
 \caption{(color online). Illustration of the double-fishnet structure with a square
  unit cell of side-length $p$ highlighted.
  The two perforated silver films are embedded
  in a dielectric host material which holds the dye molecules (translucent).
  Dimensions are given in the text. Pump  (red dashed line) and probe
  (blue solid line) pulses illustrate the pump-probe configuration
  with the electric field polarized along the $x$ direction.}
 \label{fig:figure1}
\end{figure}

The considered structure is an optical double-fishnet metamaterial 
\cite{Sivan_2009,Zhang_2005,Zhang_2005_2} with a square periodicity of $p=280\,\textrm{nm}$,
perforated with rectangular holes of sides $a_{x}=120\,\textrm{nm}$
and $a_{y}=80\,\textrm{nm}$ (see Fig.~\ref{fig:figure1}).
The additional geometrical parameters are $h_{m}=40\,\textrm{nm}$,
$h_{d}=60\,\textrm{nm}$, and $h_{c}=60\,\textrm{nm}$. This type
of NRI metamaterial, which has been the topic of intense research (see,
e.g., \cite{Sivan_2009,Zhang_2005,Zhang_2005_2,Mary_2008}), exhibits low absorption compared
to other metamaterials in the optical wavelength range. Its relatively
low absorption makes it the most promising structure for complete
loss-compensation \cite{Shalaev_2010}.

We consider two configurations, passive and active. In the passive
configuration two silver fishnet films are embedded inside a dielectric host
that has a real refractive index of $n_{h}=1.62$ (see Fig.~\ref{fig:figure1}).
The permittivity of silver follows a Drude model corrected by two
Lorentzian resonances to match experimental data at visible wavelengths
\cite{McMahon_2009}. In the active configuration we insert Rhodamine
800 dye molecules into the dielectric host and excite them optically 
in numerical pump-probe experiments. The chosen geometric parameters
ensure a good overlap of the metamaterial's resonant response with
the emission spectrum of the dye for an electric field polarization
along the long side $a_{x}$ of the rectangular holes (see Fig.~\ref{fig:figure1}).

In order to self-consistently calculate the gain dynamics in this
system the dye molecules are described using a semiclassical four-level
model with two optical dipole transitions 
\cite{Klaedtke_2004,Klaedtke_2006,Bohringer_2008,Bohringer_2008_2}.
This model is implemented by introducing auxiliary differential equations
for the position- and time-dependent polarization densities $\mathbf{P}_{i}$ 
and occupation densities $N_{j}$ into the three-dimensional finite-difference
time-domain (FDTD) algorithm. The time evolution of the polarization
densities for the absorption ($i=\textrm{a}$) and emission ($i=\textrm{e}$)
lines is then given by
\begin{equation}
 \frac{\partial^{2}\mathbf{P}_{i}}{\partial t^{2}}+2\Gamma_{i}
 \frac{\partial\mathbf{P}_{i}}{\partial t}+\omega_{0,i}^{2}
 \mathbf{P}_{i}=-2\omega_{i}\frac{e^{2}d_{i}^{2}}{\hbar}
 \Delta N_{i}\cdot\mathbf{E}_{\mathrm{loc}}\label{eq:pol}
\end{equation}
where $\omega_{0,i}=(\omega_{i}^{2}+\Gamma_{i}^{2})^{1/2}$
are the oscillator frequencies, $\hbar\omega_{i}$ the electronic transition
energies, $\Gamma_{i}$ the half-widths of the resonances, and $e\cdot d_{i}$
the dipole strengths.
The dye molecules embedded in the dielectric host
experience, in the Lorentz approximation, the local electric field
$\mathbf{E}_{\mathrm{loc}}=\left[(n_{h}^{2}+2)/3\right]\mathbf{E}$
and not the average electric field $\mathbf{E}$ \cite{deVries_1998}.
Saturation is accounted for by the electric field
dependence of the occupation inversions $\Delta N_{\textrm{a}}=N_{3}-N_{0}$
and $\Delta N_{\textrm{e}}=N_{2}-N_{1}$ for absorption and emission
respectively, which couple to Eq.~\eqref{eq:pol}. Their dynamics
are governed by
\begin{subequations}
 \begin{gather}
  \frac{\partial N_{3}}{\partial t}=\frac{1}{\hbar\omega_{\textrm{a}}}
   \left(\frac{\partial\mathbf{P}_{\textrm{a}}}{\partial t}+
   \Gamma_{\textrm{a}}\mathbf{P}_{\textrm{a}}\right)\cdot
   \mathbf{E}_{\mathrm{loc}}-\frac{N_{3}}{\tau_{32}},\\
  \frac{\partial N_{2}}{\partial t}=\frac{N_{3}}{\tau_{32}}
   +\frac{1}{\hbar\omega_{\textrm{e}}}\left(\frac{\partial
   \mathbf{P}_{\textrm{e}}}{\partial t}+\Gamma_{\textrm{e}}
   \mathbf{P}_{\textrm{e}}\right)\cdot\mathbf{E}_{\mathrm{loc}}
   -\frac{N_{2}}{\tau_{21}},\\
  \frac{\partial N_{1}}{\partial t}=\frac{N_{2}}{\tau_{21}}
   -\frac{1}{\hbar\omega_{\textrm{e}}}\left(\frac{\partial
   \mathbf{P}_{\textrm{e}}}{\partial t}+\Gamma_{\textrm{e}}
   \mathbf{P}_{\textrm{e}}\right)\cdot\mathbf{E}_{\mathrm{loc}}
   -\frac{N_{1}}{\tau_{10}},\\
  \frac{\partial N_{0}}{\partial t}=\frac{N_{1}}{\tau_{10}}
   -\frac{1}{\hbar\omega_{\textrm{a}}}\left(\frac{\partial
   \mathbf{P}_{\textrm{a}}}{\partial t}+\Gamma_{\textrm{a}}
   \mathbf{P}_{\textrm{a}}\right)\cdot\mathbf{E}_{\mathrm{loc}}.
 \end{gather}
\end{subequations}
Nonradiative decay of the occupation densities
is quantified by the lifetimes $\tau_{jk}$. The $\Gamma_{i}\mathbf{P}_{i}$-terms
stem from the transformation from complex- to real-valued polarizations
(\cite{Bohringer_2008}, p.~174).

Absorption and emission cross-sections, taken from experimental data,
are used to calculate the dipole length
$d_{i}$ via $\sigma_{i}=\left(\omega_{0,i}e^{2}d_{i}^{2}/\hbar\right)/
\left(\epsilon_{0}cn_{h}\Gamma_{i}\right)$,
with $\epsilon_{0}$ being the vacuum permittivity and $c$ the vacuum
speed of light. The parameters for the four-level system are chosen
as follows (cf. \cite{Sperber_1988}): $\lambda_{\textrm{e}}=2\pi
c/\omega_{\textrm{e}}=710\,\textrm{nm}$, $\lambda_{\textrm{a}}=680\,
\textrm{nm}$, $\Gamma_{\textrm{e}}=\Gamma_{\textrm{a}}=1/\left(20\,
\textrm{fs}\right)$, $d_{\textrm{e}}=0.09\,\textrm{nm}$, and 
$d_{\textrm{a}}=0.1\,\textrm{nm}$; $\tau_{32}=\tau_{10}=100\,\textrm{fs}$
and $\tau_{21}=500\,\textrm{ps}$. These values correspond to
cross-sections $\sigma_{\textrm{e}}=2.43\times10^{-16}\,\textrm{cm}^{2}$
and $\sigma_{\textrm{a}}=3.14\times10^{-16}\,\textrm{cm}^{2}$. We
set the density of the dye molecules as $N=\sum_{j=0}^3 N_j=
6\times10^{18}\,\textrm{cm}^{-3}\approx10\,\textrm{mM}$ leading to a 
bulk gain coefficient of approximately $g\approx N\cdot\sigma_{
\textrm{e}}\approx1460\,\textrm{cm}^{-1}$ at full inversion. 

\begin{figure}[b]
 \begin{centering}
  \includegraphics{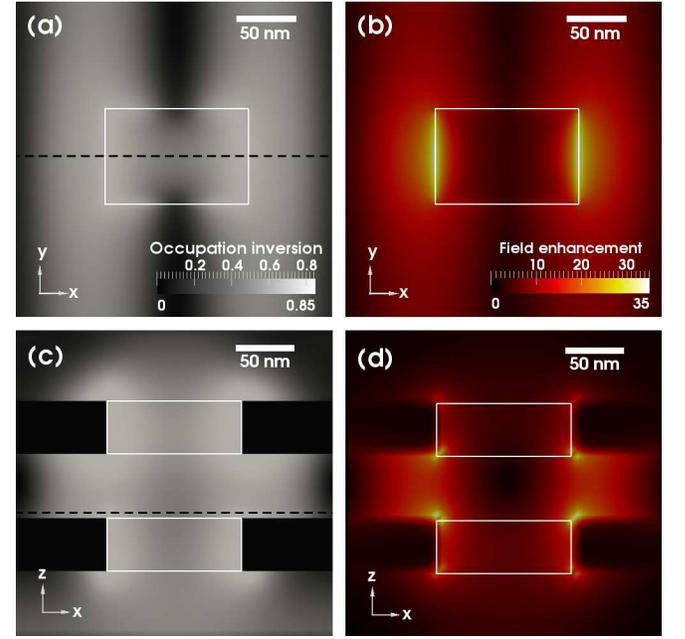} 
 \end{centering}
 \caption{(color online). (a) Snapshot of the occupation inversion $\Delta N_\textrm{e}$
  in a plane $5 \, \textrm{nm}$ below the upper silver fishnet film just
  before probing and (b) the electric field enhancement at $710\,\textrm{nm}$
  in the same plane;
  both for a pump-field amplitude of $2.0\,\textrm{kV/cm}$. The silver
  film perforation is indicated by a white rectangle. (c) and (d) show
  the same physical quantities in a plane given by the propagation direction
  and the long side of the perforation. The dashed black lines in (a)
  and (c) highlight the positions of the intersection with the other plane, respectively.
 \label{fig:figure2}}
\end{figure}

In order to study the active configuration we first pump the dye
molecules with a short, intense pulse of duration $2\,\textrm{ps}$.
After a short delay of $7\,\textrm{ps}$  we probe the structure with
a weak broadband pulse of duration $12\,\textrm{fs}$. Figures~\ref{fig:figure2}(a)
and (c) show a snapshot of the spatial distribution of the occupation
inversion generated by the pump pulse at two perpendicular planes
inside the unit cell of the active metamaterial. The effective gain
coefficient in the structure can be maximized with a good matching
between the spatial distribution of the inversion and that of the plasmon-enhanced
electric field amplitude at the emission wavelength $\lambda_{\textrm{e}}$
[Figs.~\ref{fig:figure2}(b) and (d)]. Indeed, we see from Fig.~\ref{fig:figure2}
that such a matching is achieved when the pump and the probe have
the same electric field polarization.

We remark that the considered planar structure has a low cavity $Q$-factor
($Q<50$). In addition, the modal volume, which is directly associated
with the feedback from metallic interfaces, is orders of magnitude
larger than in the case of nanospheres used for ``spasing'' \cite{Noginov_2009}. 
Therefore lasing is not expected in this structure. This is also supported by
the fact that in our simulations we do not observe any appreciable
depletion of the occupation inversion by the probe pulse. 
Further, on the short ps time-scale under investigation, which is 
3 orders of magnitude shorter than the free-space spontaneous 
emission lifetime of the dye, amplified spontaneous emission is not
expected to play a significant role \cite{Gehrig_2003}.

\begin{figure}
 \begin{centering}
  \includegraphics{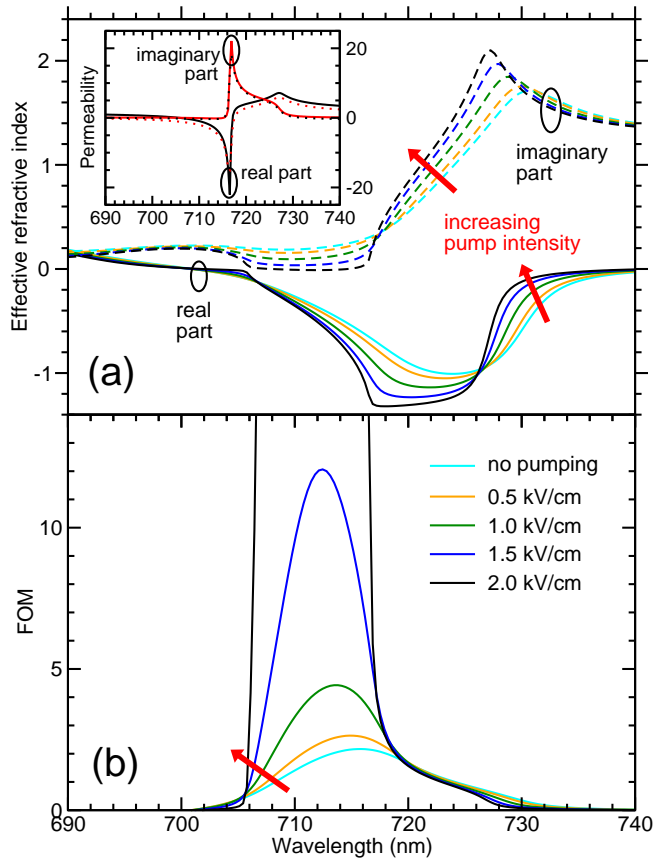} 
 \end{centering}
 \caption{(color online). (a) Real and imaginary part of the retrieved effective refractive indices
  of the double-fishnet structure for different pump amplitudes. The peak electric field
  amplitude of the pump increases in steps of $0.5\,\textrm{kV/cm}$
  from no pumping (cyan line, lightest) to a maximum of $2.0\,\textrm{kV/cm}$
  (black line, darkest). The inset shows the real and imaginary part
  of the effective permeability (black and red line, respectively) and
  the result of the Kramers-Kronig relation (black and red dotted lines)
  for the highest peak electric field amplitude of $2.0\,\textrm{kV/cm}$.
  (b) The figures-of-merit (FOM) for the same pumping amplitudes.
 \label{fig:figure3}}
\end{figure}

We use the standard retrieval method \cite{Smith_2002} to extract
the effective refractive index $n=n'+in''$, first, of the passive
configuration. Note that the metamaterial structure considered in
this work is surrounded by air above and below the dielectric host; i.e.,
it is deliberately not placed on a thin substrate, in order to be
symmetric and nonbianisotropic \cite{Kriegler_2009}. The spectral
variation of $n'$ ($n''$) in the passive structure
is similar to that shown by the cyan solid (dashed)
line in Fig.~\ref{fig:figure3}(a) (corresponding to the metamaterial
that has dye molecules included but is not pumped). We find that in
this passive case the figure of merit $\textrm{FOM}=-n'(\lambda)/n''(\lambda)$
has a maximum value of $2.7$ at $713\,\textrm{nm}$. 

\begin{figure}
 \begin{centering}
  \includegraphics[width=.99\columnwidth]{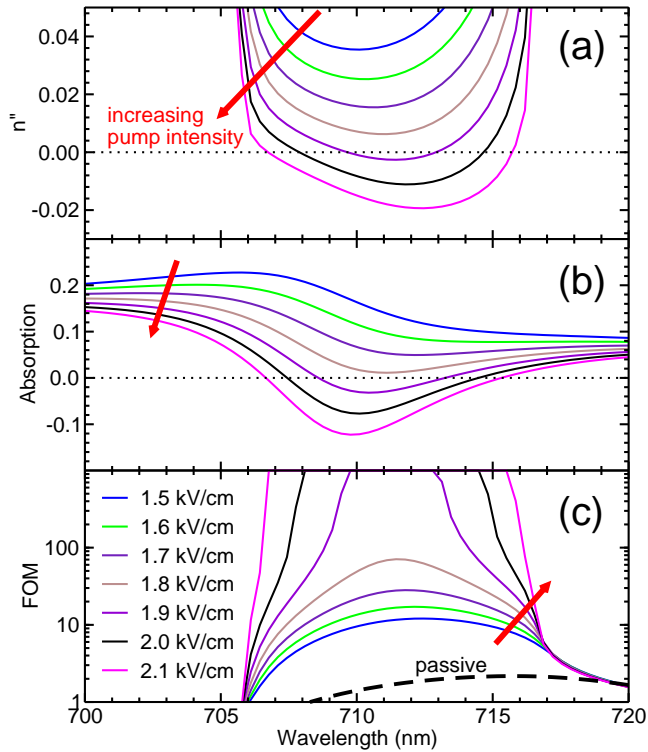} 
 \end{centering}
 \caption{(color online). Detailed view of (a) the imaginary part of the retrieved effective refractive
  indices $n''$, (b) the absorption, and (c) the figures of merit (FOM) for peak
  pump-field amplitudes close to and above compensation between $1.5$
  and $2.1\,\textrm{kV/cm}$ in steps of $0.1\,\textrm{kV/cm}$.
 \label{fig:figure4}}
\end{figure}

Next, we study the active metamaterial configuration and,
in particular, the effect that an increase of the pump intensity has on the obtained
FOM. In the calculations we ensure that the gain system
is probed within the linear regime where the standard retrieval method
can be applied.
Figure~\ref{fig:figure3}(a)
shows the real and imaginary parts of the retrieved effective refractive
indices for progressively increased peak pump-field amplitudes.
We see that increasing the pump intensity, and therefore the gain
available for the probe, leads to a decrease in the imaginary part
of the refractive index in the region around the maximum emission
cross-section of the dye ($710\,\textrm{nm}$) indicating
reduced optical losses. The stronger resonance that arises from
the gradually intensified pump also reduces the real part of the
refractive index towards more negative values. Increasing the pump pulse
amplitude from $0.5\,\textrm{kV/cm}$ to $1.5\,\textrm{kV/cm}$ 
improves and blue-shifts the maximum of the FOM from a value of $3$ at 
$716\,\textrm{nm}$ to a value of $12$ at $712\,\textrm{nm}$ (for the 
higher amplitude). An even higher amplitude of the pump-pulse of $2\,\textrm{kV/cm}$
results in a wavelength region (approx. $706-714\,\textrm{nm}$)
where the losses in the metamaterial are completely compensated.
In this region, the active metamaterial exhibits a negative
absorption [cf. Figure~\ref{fig:figure4}(b)] and both the real and
imaginary part of the refractive index become simultaneously
negative. Note from Fig.~\ref{fig:figure3}(b) that in this case
the FOM diverges at the two wavelength points bounding the negative-absorption
(amplification) region owing to $n''$ becoming exactly zero at these
two wavelengths. 

To further verify the causal nature of the obtained effective parameters
we use the method of \cite{Cook_2009} to calculate, based on the
Kramers-Kronig relations, the real (imaginary) part of the effective
permeability from the imaginary (real) part of the numerically retrieved \cite{Smith_2002}
effective permeability. An example of such a calculation for a pump amplitude of $2\,\textrm{kV/cm}$
(corresponding to the negative-absorption regime) is shown in the
inset of Fig.~\ref{fig:figure3}(a). The excellent agreement between
the results obtained from the standard retrieval method
and the complementary Kramers-Kronig approach further confirms that 
the extracted parameters do obey causality. 

Finally, Fig.~\ref{fig:figure4}
presents a more detailed look at $n''$, the absorption coefficient,
and the FOM for pump intensities close to and above complete loss compensation.
We note that there is a critical amplitude of approximately $1.85\,\textrm{kV/cm}$
for the pump pulse beyond which the present metamaterial configuration
becomes amplifying. A further increase of the pump field up to levels
of $2.1\,\textrm{kV/cm}$ leads to a spectral broadening of the region
of negative absorption. At very large pump-field amplitudes above
$2.2\,\textrm{kV/cm}$ we observe discontinuities in the effective
refractive index (data not shown here) and the Kramers-Kronig relation
for the permeability is not obeyed any more. This occurs despite a
smooth change in amplitude and phase of both the transmission and
the reflection coefficients with no sign of gain depletion or discontinuities
in other physical quantities. Investigation of this regime will be the subject of
future work.

In conclusion, we have shown how the incorporation of a gain medium
(Rhodamine 800 dye) into the structure of a double-fishnet nonbianisotropic
metamaterial can fully compensate losses in the regime where the real
part of the metamaterial's effective refractive index is negative.
In this spectral range the imaginary part of the refractive index
of the metamaterial becomes negative, too. We believe that this work
could guide experimental efforts towards lossless and amplifying metamaterials,
which offer access to exciting photonic applications. 

We gratefully acknowledge financial support provided by the EPSRC
and the Royal Academy of Engineering.


%

\end{document}